# An In-Ear Sleep Modulation SoC Featuring a CNN–LSTM Accelerator with Long-Kernel Memory Reuse and Runtime Dynamic Quantization

Yaqian Xu[1,*], Xinyu Chen[1,*], Yuhan Hou[1], Jianxiong Xu[1], Hao You[1], Yu Huang[1], Bowen Liu[1], Ashley Hung[1], Andrew G. Richardson[2], Roman Genov[1], and Xilin Liu[1]

[1]University of Toronto, Toronto, ON, Canada, [2]University of Pennsylvania, Philadelphia, PA, USA, [*]Equal contribution authors

## Abstract

This work presents a 65 nm SoC for closed-loop in-ear sleep modulation. The SoC performs low-noise in-ear EEG sensing, CNN–LSTM-based sleep stage classification, and phase-specific auditory stimulation under 1 mW. A dedicated accelerator is developed, featuring a memory reuse technique to efficiently process long kernels required for low-frequency EEG feature extraction. A runtime dynamic quantization is incorporated, reducing model size by 3.88× with only a 0.28% loss in accuracy. A sensitivity of 98.3% is achieved for deep sleep detection, enabling effective sleep modulation.

## Introduction

Sleep is essential for physiological health, yet about 40% of adults experience sleep impairments that negatively affect cognitive functions, cardiovascular health, and contribute to neurodegenerative diseases such as Alzheimer's disease [1]. Deep sleep (i.e., N3 stage) and its hallmark slow-wave activity (SWA) around 1 Hz are crucial for memory consolidation and cognitive function [1]. SWA disruptions can lead to cognitive decline, whereas enhancing it improves cognitive outcomes. Although exercise and meditation can influence SWA, their effects are inconsistent. More reliable improvements have been achieved using SWA phase-specific auditory stimulation during N3, which can be detected from EEG (including in-ear EEG) by deep learning (DL) models [2]. However, existing solutions rely on costly and time-consuming sleep laboratory visits, hindering long-term adoption. As a result, wearable home-use sleep modulation devices are highly desirable.

This work presents an in-ear SoC for closed-loop auditory sleep modulation (Fig. 1). Unlike conventional EEG headsets, the proposed SoC records EEG from the ear canal, allowing the entire device to fit comfortably within the ear, similar to wireless earbuds, while simultaneously delivering auditory stimulation. Central to this SoC is an accelerator optimized for a CNN–LSTM model for sleep stage classification, achieving 98.3% sensitivity in detecting N3 on the public MASS SS3 dataset [3]. The accelerator introduces a serpentine dataflow to enable local memory reuse to efficiently process the long CNN kernels required for sleep feature extraction. In addition, it supports runtime dynamic mixed-precision quantization to mitigate accuracy loss while reducing computational cost. The accelerator is integrated with a low-noise in-ear EEG front-end featuring an input-referred noise of 0.65 μV, ΔΣ ADCs achieving a Schreier FoM of 172 dB, a mixed-signal SWA phase detector, a pink-noise generator with an audio amplifier, and a wireless module. Upon detection of N3 sleep, the SoC delivers pink-noise stimulation aligned to a desired phase of the SWA cycle, enabling closed-loop sleep modulation.

## Implementation and Measurements

The DL accelerator consists of a processing element (PE) array with 2 Super PEs and 6 regular PEs (Fig. 2). All PEs support mixed-precision MAC operations, while the Super PEs incorporate advanced activations and quantization. CNN–LSTM models can effectively classify sleep stages using long temporal kernels [2]. However, such kernels are inefficiently supported by conventional accelerators that are predominantly optimized for the short CNN kernels commonly used in computer vision and natural language processing tasks. We developed a specialized dataflow that splits kernels by stride and maps activations across the PE array in a serpentine pattern (Fig. 3a). After each cycle, PEs fetch neighboring activations and partial sums using nearby registers, reducing memory access by 2.25× and 6.2× compared to weight- and output-stationary dataflows, respectively (Fig. 3b).

Conventional static quantization reduces computational cost but degrades performance for EEG-based models with large signal dynamic ranges over long time windows and fails to generalize well across inter-subject variability. To address this, we developed a runtime dynamic quantization scheme in mixed-precision, where an on-chip histogram observer tracks the data distribution and updates the quantization parameters (Fig. 4). Inputs and inter-layer activations are quantized to 2b, 4b, or 8b on the fly (weights are pre-quantized during training). Both static and dynamic quantization are used to preserve model performance while minimizing runtime computational overhead. For standardized benchmarking, the performance of the model is evaluated on the public MASS dataset, which can be adapted to in-ear EEG through transfer learning. Testing is strictly on subjects unseen during training. Compared to our floating-point baseline, static 8b quantization suffers a 5.01% accuracy drop, whereas the hybrid approach achieves a 3.88× model size reduction with only 0.28% degradation (Fig. 5a).

LSTM operations are mapped to the PE array using an input-stationary dataflow with sequential gate executions (Fig. 5b). Activation functions are implemented using LUT-free piecewise-linear approximations. To efficiently support the various bit-widths used across different layers and kernels, we developed a bit-flexible multiplier (Fig. 6). Prior designs often fail to balance low-bit flexibility with high-bit efficiency. We combine sub-multipliers to support 8 parallel 2b, 4 parallel 4b, 2 parallel 8b, or a single 16b multiplication, optimizing the balance between speed (via parallelism), area, and power. Area savings are demonstrated through comparison with a state-of-the-art multiplier design (Fig. 6). The accelerator delivers an average inference power of 0.518 mW with a peak efficiency of 635 GOPS/W (Fig. 7), while supporting unique features with competitive power and area efficiency (Table I).

The SoC integrates a chopper-stabilized EEG amplifier achieving an input-referred noise of 0.65 μV with a NEF of 2.98 (Fig. 8a), and 2nd-order ΔΣ ADCs achieving 89 dB SNDR and a Schreier FoM of 172 dB (Fig. 8b). In-ear EEG was collected using 3D-printed earpieces and subsequently tested with the chip for sleep stage classification and closed-loop stimulation. Fig. 8c shows the characteristic time- and frequency-domain features of N2 (light sleep) and N3, confirming the classification results. SWA phase detection is illustrated in Fig. 9a. Upon detection of N3, a programmable delay aligns 50 ms pink-noise bursts to a desired SWA phase, driving a class-D audio amplifier. Fig. 9b shows the auditory stimulation bursts triggered in the closed-loop configuration, and Fig. 9c confirms the target phase distribution. A wireless module is integrated to upload EEG for validation and model training, achieving 2 Mbps with a TX power of -15 dBm over 15 cm using OOK. The SoC consumes 0.985 mW during typical operation. Fig. 10 shows a photo of an in-ear prototype, a chip micrograph, and a summary of chip specifications and model performance (Table II).

## Conclusion

Compared with state-of-the-art neuromodulation designs (Table III), this work stands out in system integration, dynamically quantized DL accelerator, circuit performance and efficiency. As earables emerge as a key trend in next-generation wearables, the technologies developed here enable new opportunities for future ear-EEG-based applications.

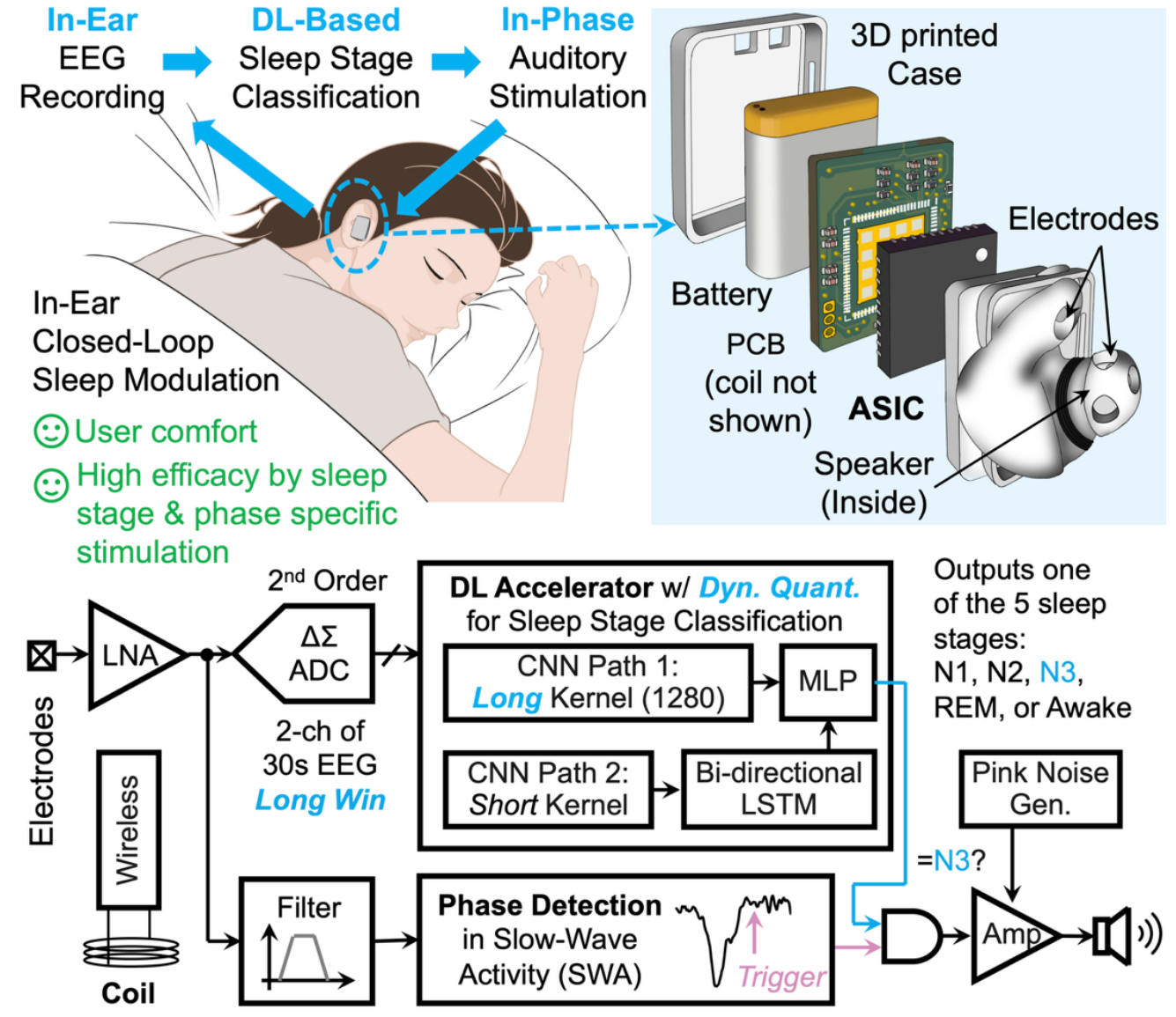

Fig. 1. Overview of the in-ear closed-loop sleep modulation device and a simplified system block diagram.

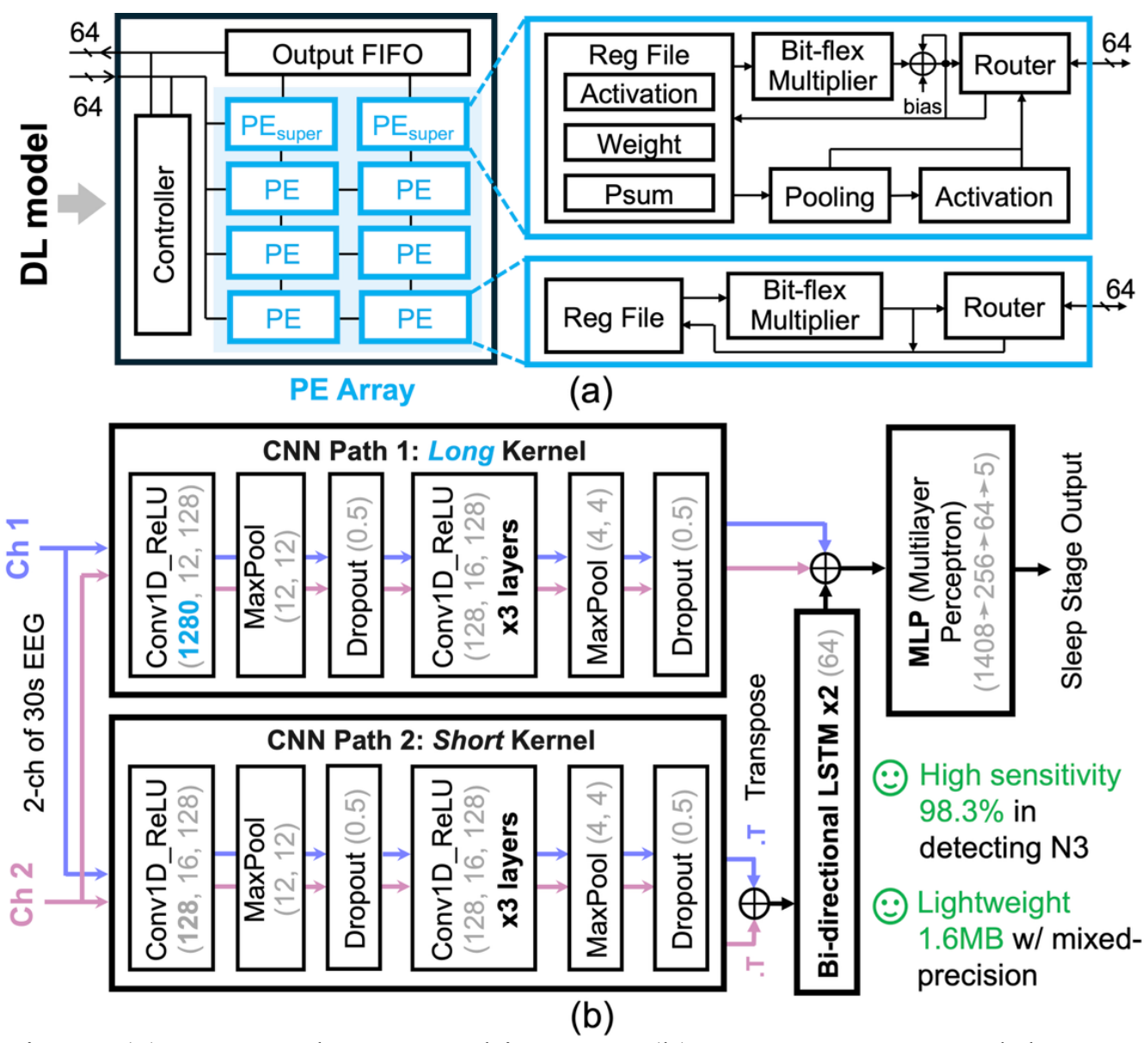

Fig. 2. (a) DL accelerator architecture. (b) CNN–LSTM model.

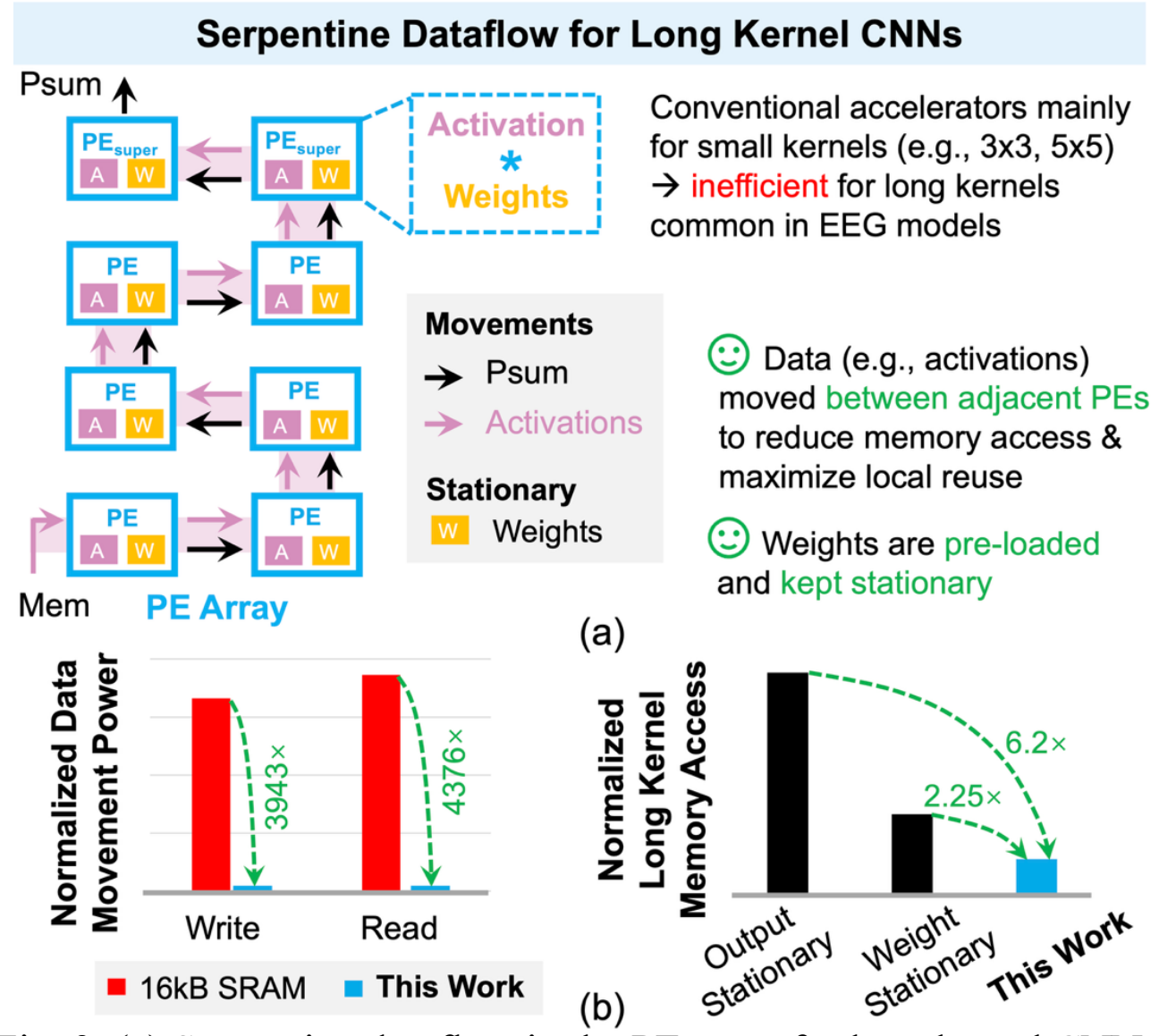

Fig. 3. (a) Serpentine dataflow in the PE array for long kernel CNNs. (b) Comparison of efficiency with conventional implementations.

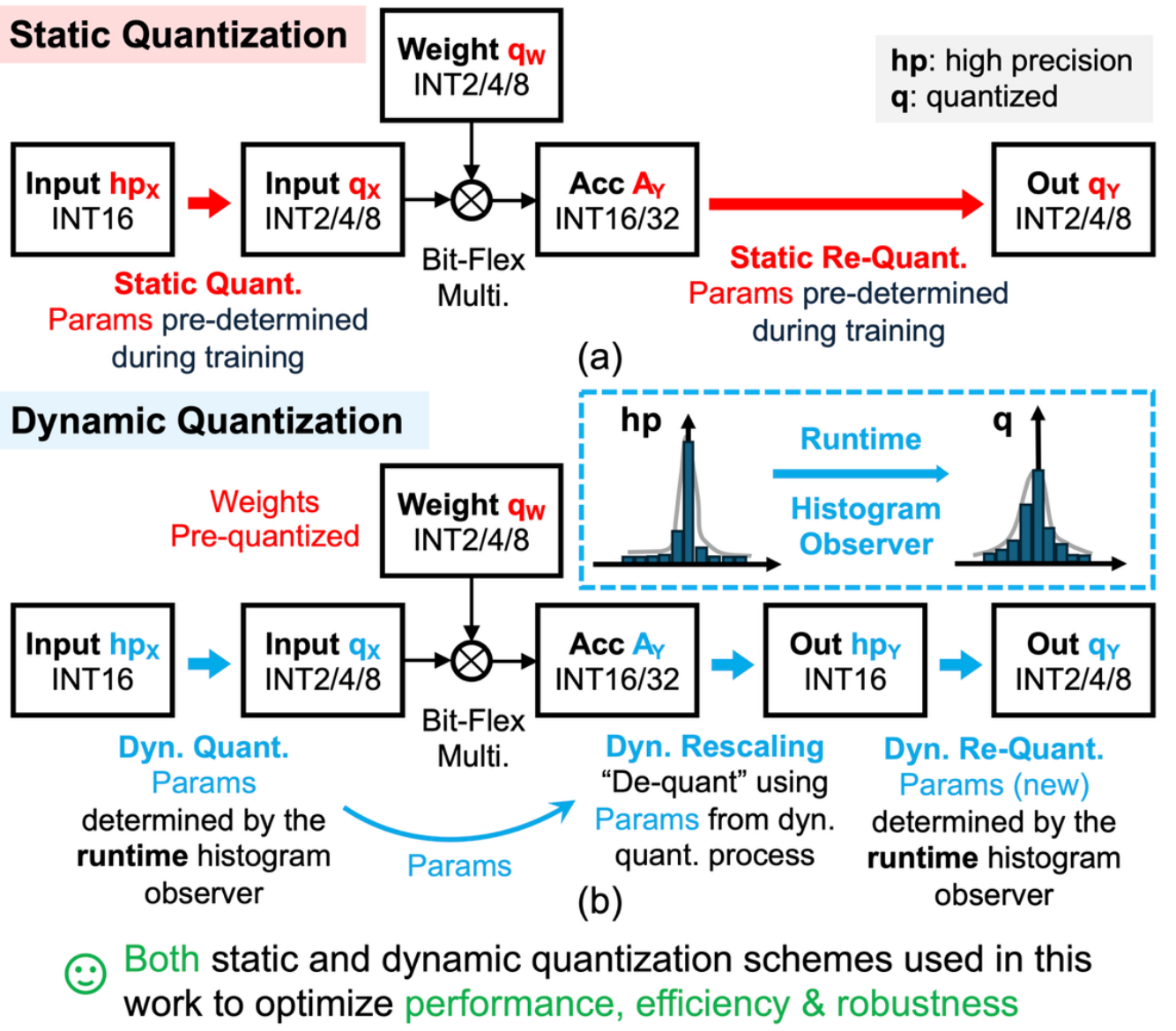

Fig. 4. (a) Static quantization and (b) dynamic quantization with runtime histogram observer.

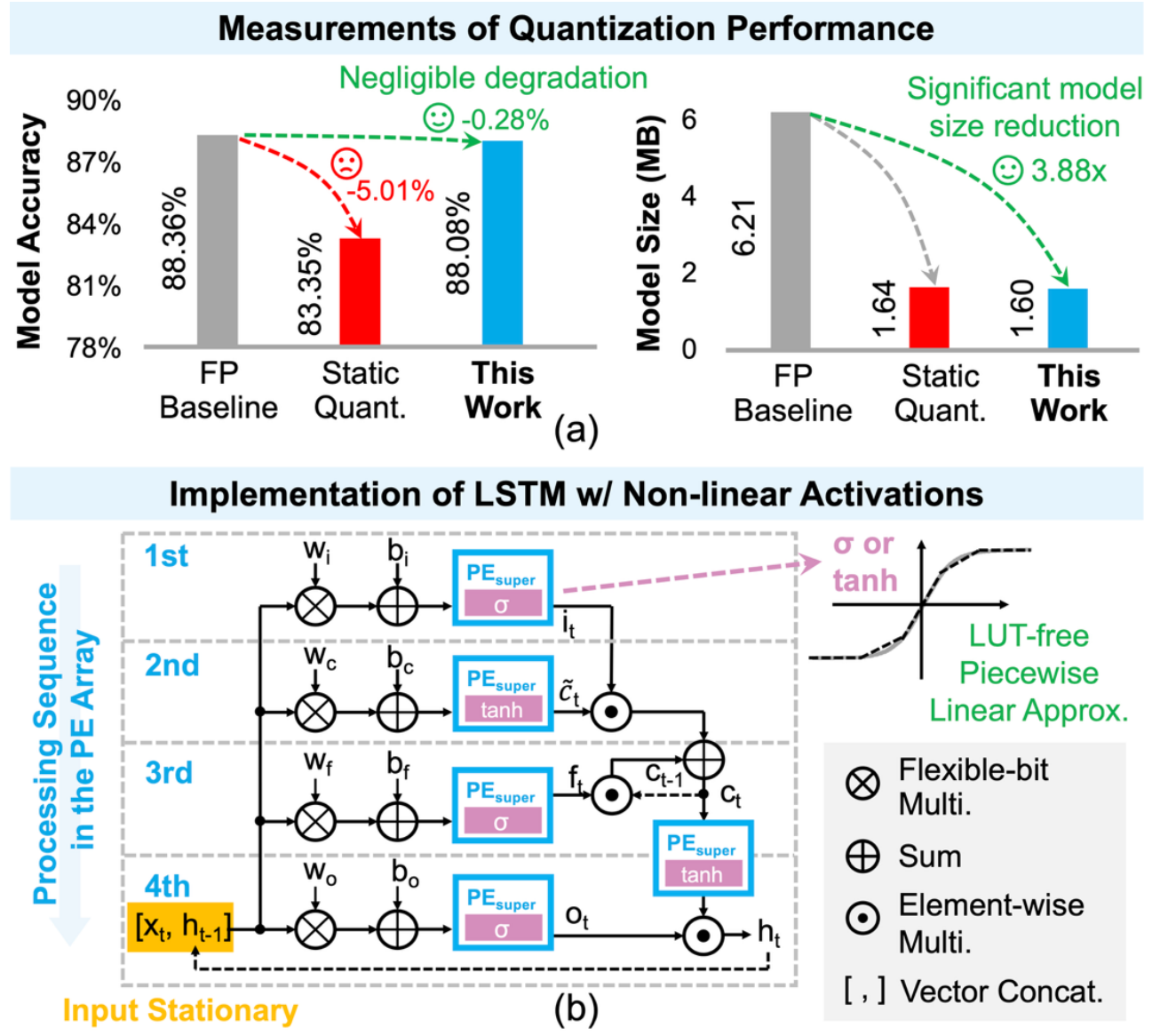

Fig. 5. (a) Model accuracy and size under different quantization schemes. (b) LSTM processing engine.

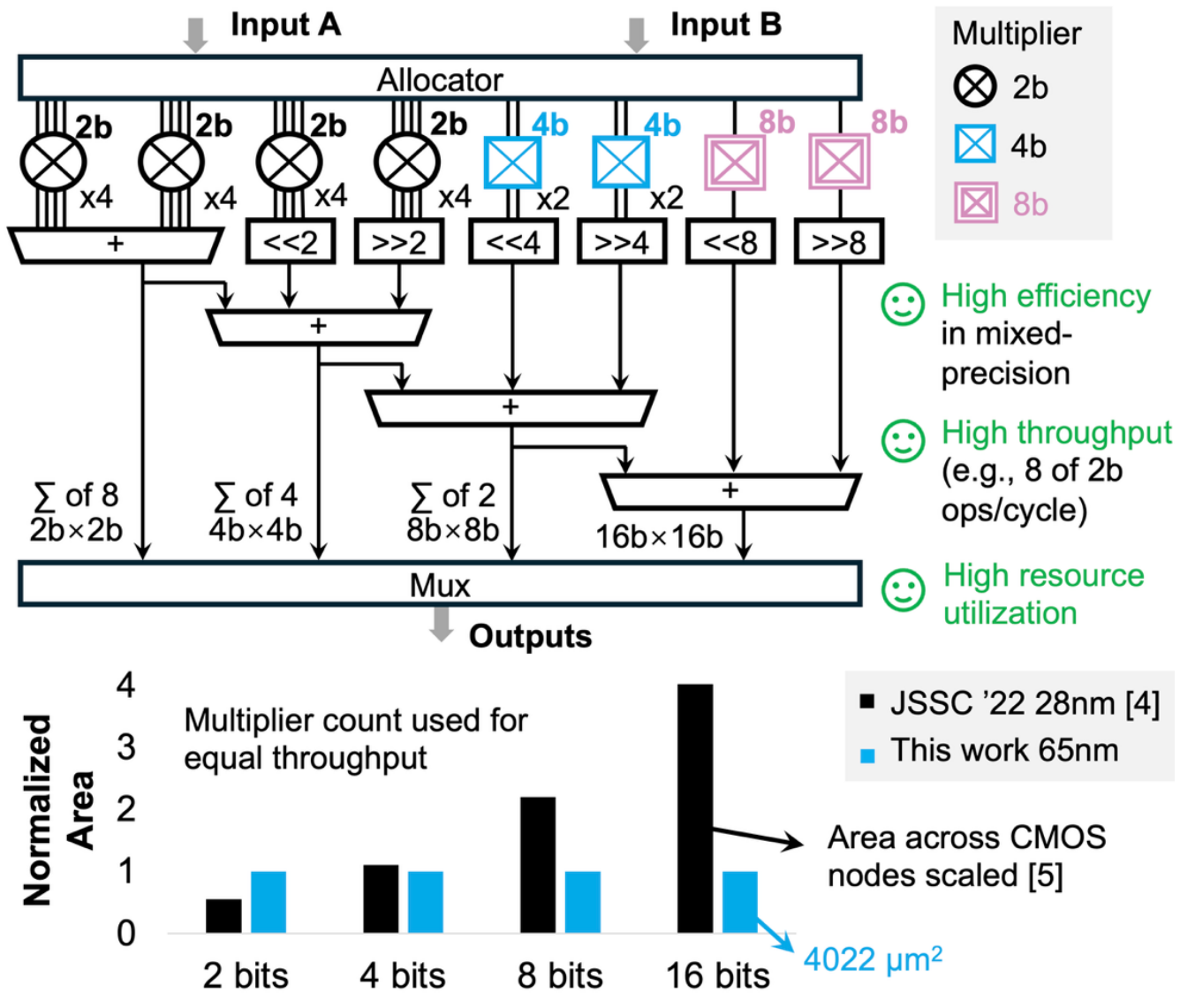

Fig. 6. The bit-flexible multiplier and area savings compared with a state-of-the-art design.

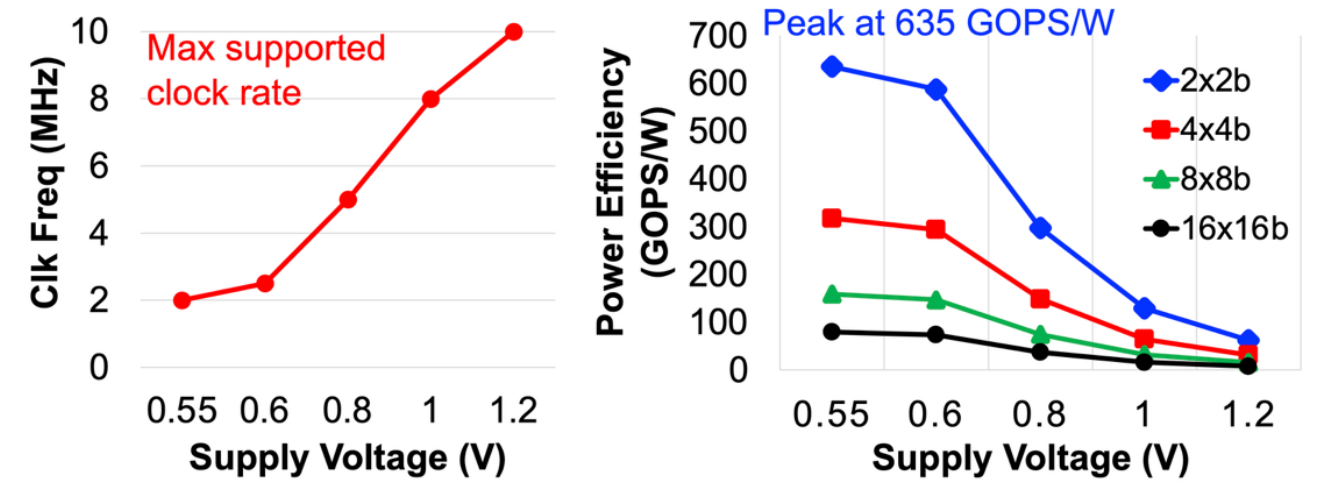

Table I Comparison with State-of-the-Art DL Accelerators

| Publication | This Work | ISSCC 25 [6] | JSSC 24 [7] | ISSCC 23 [8] | JSSC 22 [4] |
|---|---|---|---|---|---|
| CMOS | 65 nm | 40 nm | 40 nm | 28 nm | 28 nm |
| Application | Sleep stage classification | Speech | Brain pattern | Computer vision | Computer vision |
| ML Model | CNN + LSTM | RNN | CP-SVM | DNN | CNN |
| PE Area [$\mu m^2$] | PE/$PE_{super}$ 28,992/30,338 | - | 81,489 | 141,262 | 1,100 (Est.) |
| Supply [V] | 0.55-1.2 | 0.6-1.05 | 0.7-1 | 0.46-1.13 | 0.6-1 |
| DL Power [mW] | 0.24-10.48 0.518 (Typ.) | 3.5-53 | 9.68 @0.85V | 2.06-237.1 | 7.8-74 |
| Quant. Scheme | **Dynamic quant.** | - | - | Non-linear quant. | - |
| Precision | **Mixed INT2/4/8/16** | INT4/8 | - | INT1-8 | Mixed INT2/4/8 |

Fig. 7. Measured performance of the DL accelerator and comparison with state-of-the-art designs.

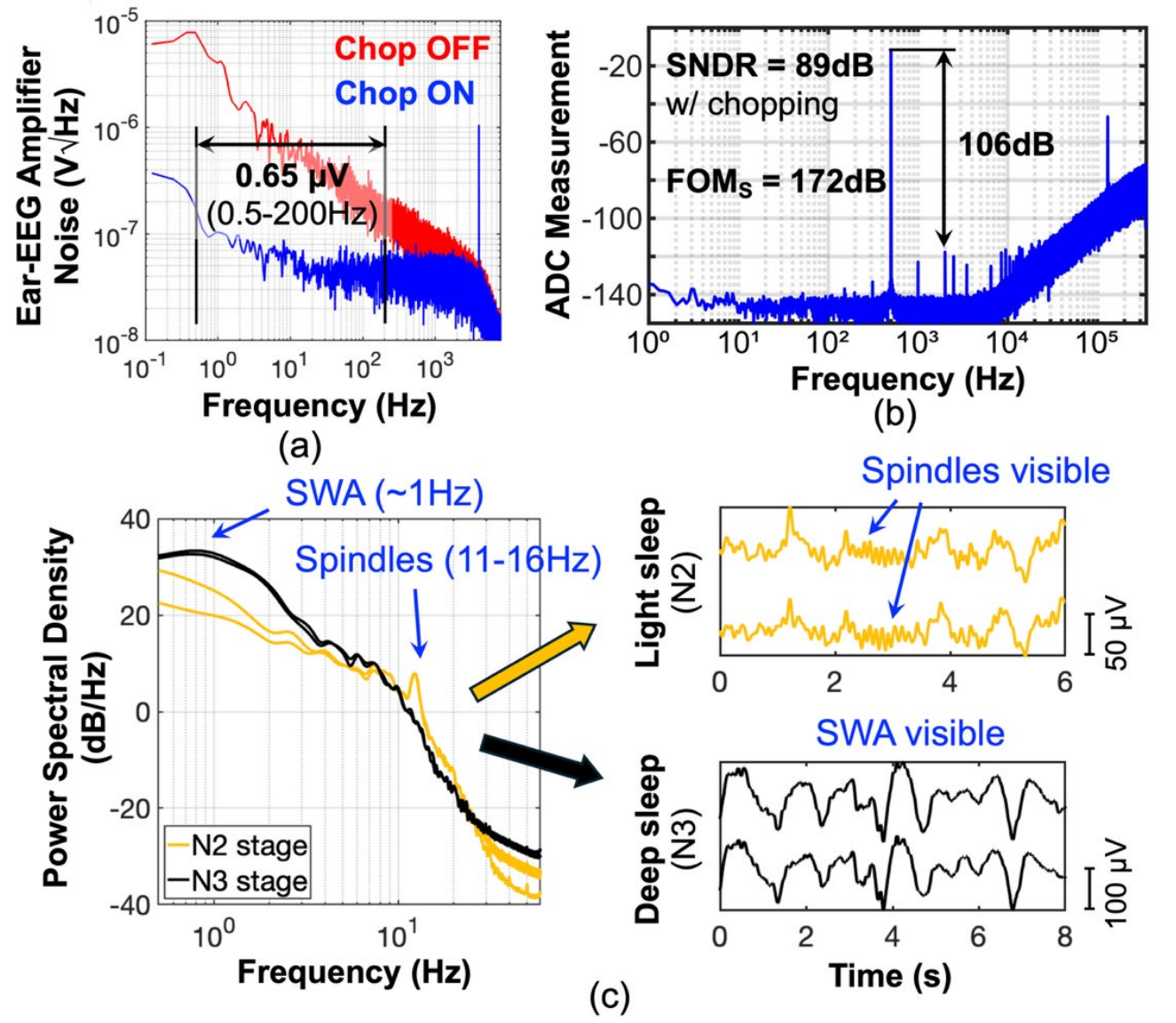

Fig. 8. Measured performance of (a) amplifier and (b) ADC. (c) In-ear EEG recordings with sleep stage classification.

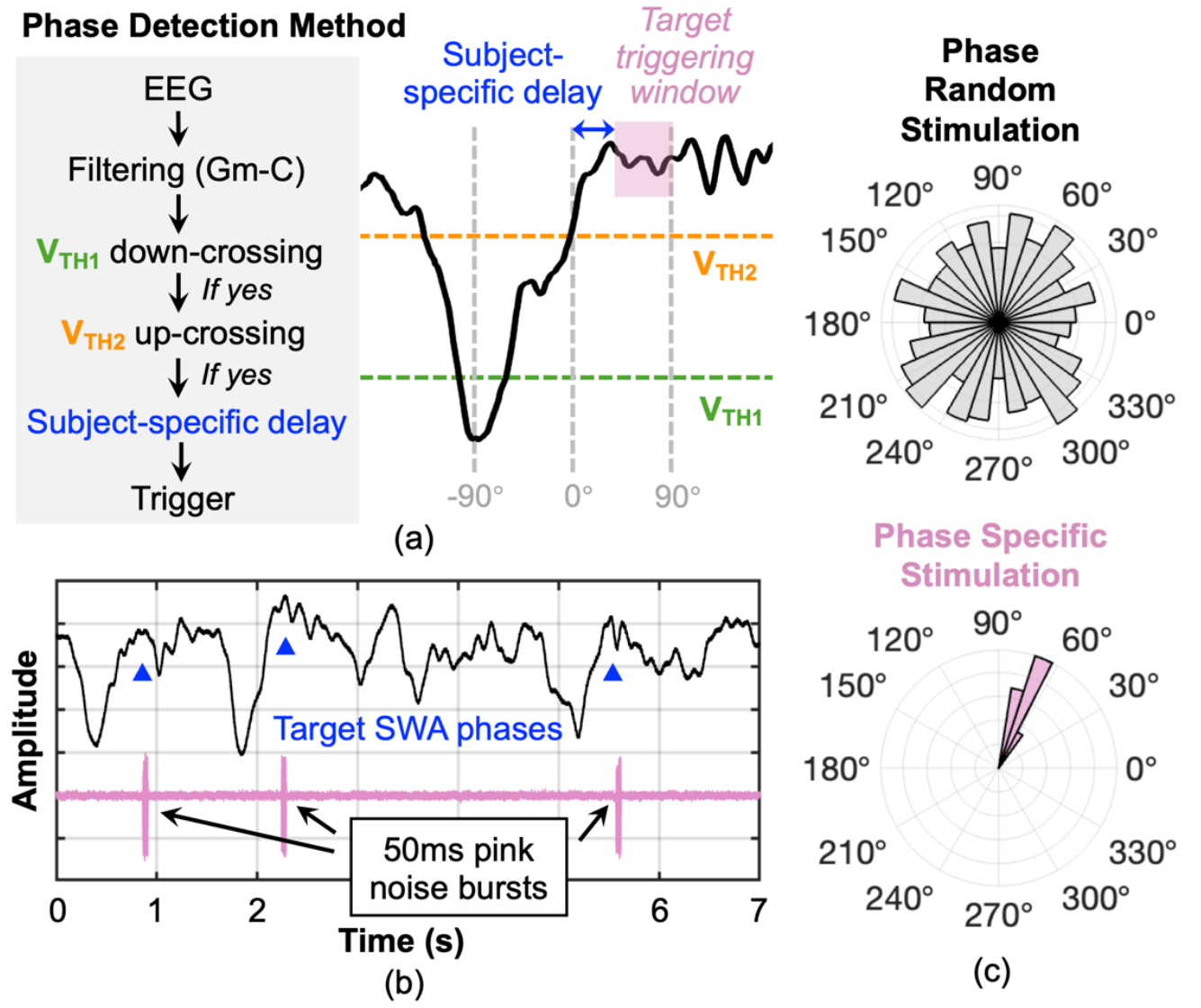

Fig. 9. (a) SWA phase detection method. (b) Measured closed-loop auditory stimulation aligned to the SWA phases. (c) Distribution of stimulation phases without and with phase-specific triggering.

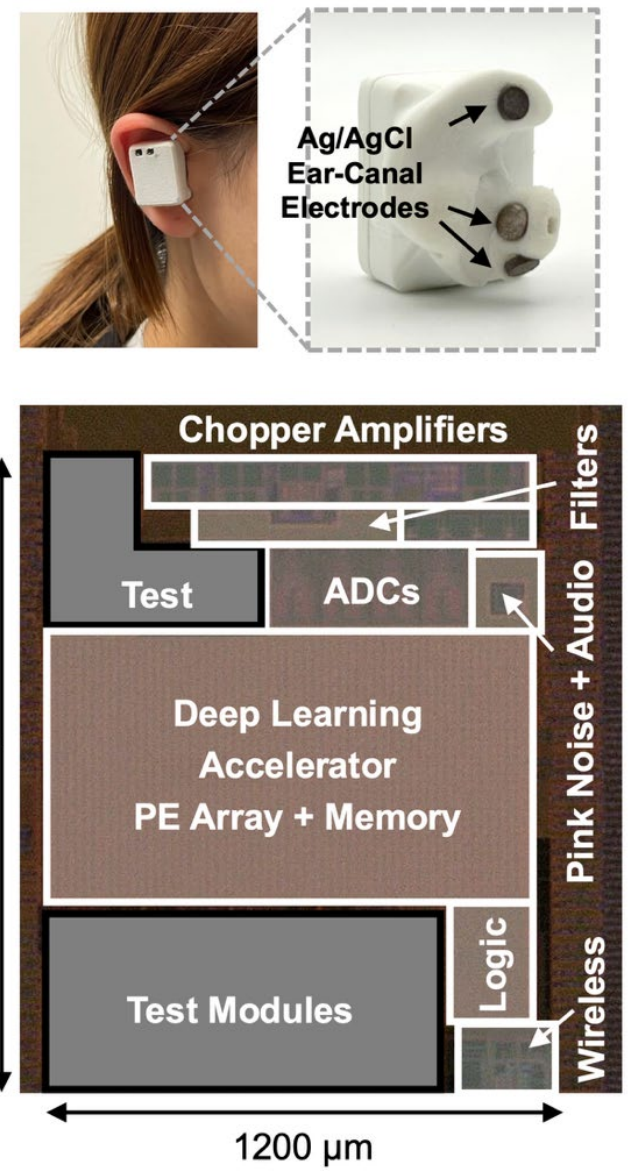

Table II Chip Specifications

| Item | Value |
|---|---|
| Silicon Area | 1.8 mm² |
| Neural Amplifier | 0.65 μV, NEF: 2.98 @4.5 μW |
| 2nd-Order ΔΣ ADC | ENOB: 14.5b FOMs: 172 dB |
| Gm-C Filter for SWA Extract | HD3: -48.4 dB @120mV input |
| Audio Amp Pwr & Efficiency | 1.3 mW (peak) >87.2% |
| Wireless TX | Tx -15 dBm @2 Mbps (>15cm) |
| DL Efficiency | 635 GOPS/W (peak) |
| DL Inference Pwr | 0.518 mW |
| Total SoC Pwr | 0.985 mW |

DL Model Performance

| N3 Detection Mode (w/ Quant.) | | |
|---|---|---|
| Deep sleep (N3) | Per-Class Sensitivity* | 98.3% |
| | Per-Class Specificity | 93.0% |
| | Per-Class Accuracy | 94.3% |
| 5-Class Avg. Accuracy | | 85.6% |
| **Balanced Mode** (w/ Quant.) | | |
| 5-Class Avg. Accuracy | | 88.1% |

*Most important for SWA sleep modulation.

Fig. 10. In-ear prototype, chip micrograph and specifications.

TABLE III. Comparison with State-of-the-art Systems

| Publication | | This Work | In-Ear or Behind-Ear Devices: Nature Comm. 24 [9] | Nature BME 23 [10] | ISSCC 23 [11] | Neuromodulation Devices: ISSCC 25 [12] | ISSCC 25 [6] |
|---|---|---|---|---|---|---|---|
| System | Process | 65 nm | 28 nm* | - | 180 nm | 65 nm | 40 nm |
| | Power Supplies [V] | 0.55-1.2/2.5 | 3/6/9/12* | - | 6-10 | 0.45/1 /2.5 | 0.6-1.05 |
| | Application | Sleep Modulation | Drowsiness Monitor | Brain States | Mental Health | Prosthesis Feedback | Speech Decoding |
| | Integrated | Yes | No | No | No | Yes | Yes |
| | Wireless | 2 Mbps | 1.96 Mbps | - | 3-18 kbps* | 200 Mbps | - |
| Neural Interface | Electrodes | Ag/AgCl dry | Au dry | Ag dry | gel wet | - | - |
| | Rec ch # | 4 ch | 11 ch | 3 ch (x2) | 2 ch | 64 ch | 128 ch |
| | AFE noise | 0.65 μV | 70 nV/√Hz | - | 0.22 μV | 5.5 μV | - |
| | AFE NEF | 2.98 | - | - | 2.68 | 2.4 | - |
| | ADC ENOB | 14.5 b | - | - | 13.8 b* | - | - |
| | ADC FOMs | 172 dB | - | - | - | 156 dB | - |
| | Stimulation | Audio | - | - | Electrical | Electrical | - |
| Processing | Processor | CNN + LSTM | LR | FBCSP | - | Custom | NN |
| | Task Specific Accuracy | 85.6%# (5-class) | 93.3% | 90.26/ 64.97% | - | 94.58% | - |
| | Sensitivity | 98.3%# | 95.6% | - | - | - | - |
| | Specificity | 93.02%# | 93% | - | - | - | - |
| | Proc. Power | 0.518 mW$ | - | - | - | 1.6 μW/ch | 3.9 mW |

*Estimated or deduced from the paper. #Performance of the N3 detection mode (summarized in Fig. 10) used for comparison. $Power of the DL accelerator during inference.